\documentclass[twocolumn,a4paper,reprint,amssymb,aps,prd,groupedaddress,superscriptaddress,amsmath,nobibnotes,showpacs,nofootinbib]{revtex4-1}

\usepackage{graphicx,bm,color,psfrag,hyperref}
\usepackage{amsfonts}
\usepackage{lipsum}
\usepackage{threeparttable,booktabs}

\newcommand{\be}{\begin{equation}}
\newcommand{\ee}{\end{equation}}

\begin{document}

\title{Screening anisotropy via energy-momentum squared gravity:\\ $\Lambda$CDM model with hidden anisotropy}
\author{\"{O}zg\"{u}r Akarsu}
\email{akarsuo@itu.edu.tr}
\affiliation{Department of Physics, Istanbul Technical University, Maslak 34469 Istanbul, Turkey}
\affiliation{Abdus Salam International Centre for Theoretical Physics, Strada Costiera 11, 34151, Trieste, Italy}

\author{John D. Barrow}
\email{J.D.Barrow@damtp.cam.ac.uk}
\affiliation{DAMTP, Centre for Mathematical Sciences, University of Cambridge, Wilberforce Road, Cambridge CB3 0WA, United Kingdom}


\author{N. Merve Uzun}
\email{nebiye.uzun@boun.edu.tr}
\affiliation{Department of Physics, Bo\u{g}azi\c{c}i University, Bebek 34342 Istanbul, Turkey}
\begin{abstract}

We construct a generalization of the standard $\Lambda$CDM model, wherein we simultaneously replace the spatially flat Robertson-Walker metric with its simplest anisotropic generalization (LRS Bianchi I metric), and couple the cold dark matter to the gravity in accordance with the energy-momentum squared gravity (EMSG) of the form $f(T_{\mu\nu}T^{\mu\nu})\propto T_{\mu\nu}T^{\mu\nu}$. These two modifications---namely, two new stiff fluid-like terms of different nature---can mutually cancel out, i.e., the shear scalar can be screened completely, and reproduce mathematically exactly the same Friedmann equation of the standard $\Lambda$CDM model. This evades the BBN limits on the anisotropy, and thereby provides an opportunity to manipulate the cosmic microwave background quadrupole temperature fluctuation at the desired amount. We further discuss the consequences of the model on the very early times and far future of the Universe. This study presents also an example of that the EMSG of the form $f(T_{\mu\nu}T^{\mu\nu})\propto T_{\mu\nu}T^{\mu\nu}$, as well as similar type other constructions, is not necessarily relevant only to very early Universe but may even be considered in the context of a major problem of the current cosmology related to the present-day Universe, the so-called $H_0$ problem.

\begin{center}
    \textit{Dedicated to the memory of Professor John David Barrow}
\end{center}

\end{abstract}

\maketitle

\section{Introduction}

The standard $\Lambda$CDM (Lambda cold dark matter) model has begun to be seen, with an increasing consensus, as an approximation to a more  realistic model that still needs to be fully understood \cite{DiValentino:2020vhf}. As it is in good agreement with most of the currently available data \cite{Riess:1998cb,Alam:2016hwk,Abbott:2017wau,Aghanim:2018eyx}, the deviations from the standard $\Lambda$CDM are not expected to be too drastic from the phenomenological point of view, even if they can be conceptually very different. Indeed, the recent theoretical (e.g., de Sitter swampland conjecture \cite{Obied:2018sgi,Agrawal:2018own,Colgain:2018wgk,Heisenberg:2018yae,Akrami:2018ylq,Raveri:2018ddi,Cicoli:2018kdo,Colgain:2019joh}) and observational (e.g., persistent tensions among some existing datasets \cite{Verde:2019ivm,Riess:2019cxk,Freedman:2019jwv,Aubourg:2014yra,Zhao:2017cud,Bullock:2017xww,tension02,Mortsell:2018mfj,Dutta:2018vmq,Vagnozzi:2019ezj,Handley:2019tkm,DiValentino:2019qzk,Akarsu:2019hmw,DiValentino:2020hov})  developments, along with the notoriously challenging theoretical issues related to $\Lambda$ \cite{Weinberg:1988cp,Peebles:2002gy}, suggest that accomplishment of a successful extension of the standard $\Lambda$CDM would not be a straightforward task. Its extensions, so far, mostly focus on replacing either $\Lambda$ (the positive cosmological constant) with a dynamical dark energy or the general relativity (GR) with a modified gravity theory \cite{Copeland:2006wr,Clifton:2011jh,DeFelice:2010aj,Capozziello:2011et,Nojiri:2010wj}. In fact, there is another option that has not been emphasized much; replacing the spatially maximally symmetric and flat Robertson-Walker (RW) metric assumption of the model with a more generic metric, e.g., with an anisotropic metric, which typically results in a dynamical geometrical modification (likewise the spatial curvature) in the usual Friedmann equation of the standard $\Lambda$CDM, the shear scalar---a measure of the anisotropic expansion. The spatially flat RW background assumption has conventionally been justified via the standard inflationary scenarios employing canonical scalar fields \cite{Starobinsky:1980te,Guth:1980zm,Linde:1981mu,Albrecht:1982wi}, wherein the space dynamically flattens and very efficiently isotropizes (cosmic no-hair theorem \cite{Wald:1983ky,Starobinsky:1982mr}). Allowing anisotropic expansion factors---while retaining isotropic spatial curvature---leads to a generalized Friedmann equation bringing in average Hubble parameter along with a shear scalar \cite{Collins,Collins:1972tf,Ellis:1998ct,GEllisBook} mimicking the stiff fluid (described by an equation of state of the form $p=\rho$ \cite{zel61,Barrow78}) and hence diluting faster than any other physical source (for which $p=\rho$ is the causality limit \cite{GEllisBook}) as the Universe expands. The stiff fluid-like shear scalar is typical for general relativistic anisotropic universes with isotropic spatial curvature filled only with isotropic perfect fluids with no peculiar velocities \cite{GEllisBook}. Hence, it is not expected there to be an anisotropic expansion at measurable levels in the observable Universe. Nevertheless, the interest in anisotropic cosmologies has never been ceased, as, for instance, deviations from the stiff fluid-like shear scalar might imply the necessity of replacing $\Lambda$ with anisotropic stresses that excludes the most common dark energy models such as the minimally coupled scalar fields. See \cite{Barrow:1997sy} for a list of well known anisotropic stresses (vector fields, spatial curvature anisotropies etc.) and their effects on the expansion anisotropy/shear scalar. This interest has frequently been reinforced by some new  observations. See, for instance, Refs. \cite{Bennett11,Ade:2013kta,Schwarz:2015cma,Akrami:2019bkn,Wilczynska:2020rxx,Migkas:2020fza} and references therein, for hints of unexpected features in the cosmic microwave background (CMB) data from the WMAP and Planck missions and in other types of independent cosmological data. And, Refs. \cite{Koivisto:2005mm,Campanelli:2006vb,Koivisto:2007bp,Rodrigues:2007ny,Koivisto:2008xf,Campanelli:2007qn,Campanelli2,Campanelli:2009tk,Cea:2019gnu,Akarsu:2020pka} suggesting that the lack of quadrupole moment in the CMB temperature angular power spectrum \cite{Bennett11,Ade:2013kta,Schwarz:2015cma,Akrami:2019bkn} can be addressed by anisotropic expansion driven, well after the matter-radiation decoupling, by anisotropic dark energy (see also \cite{Battye:2006mb,Koivisto:2008ig,Cooray:2008qn,Akarsu:2013dva,Koivisto:2014gia,Heisenberg:2016wtr,Yang:2018ubt}, and, for constraints on such models, \cite{Mota:2007sz,appleby10,Appleby:2012as,Amendola:2013qna}). Seeking possible significant deviations from isotropic expansion occupies an important place in the upcoming projects such as the Euclid mission \cite{Amendola:2016saw}, as it can be very illuminating to the nature of dark energy---namely, generically, modified gravity theories induce nonzero anisotropic stresses that lead to characteristic modifications on the dynamics of the shear scalar, see, e.g., \cite{Pimentel89,Madsen88,Faraoni:2018qdr,Akarsu:2019pvi}. All these works focus on the idea of relaxing the limits upon the anisotropic expansion by making the shear scalar less stiff, by replacing either $\Lambda$ with an anisotropic dark energy model or GR by a modified gravity theory model that can induce an anisotropic dark energy. Through such setups, the limits obtained from big bang nucleosynthesis (BBN) can be weakened considerably with respect to the ones imposed by the CMB \cite{Barrow:1997sy}. However, the Friedmann equation, say $H(z)$, in such models in general deviates from that of the $\Lambda$CDM model because of both the replacement of $\Lambda$ with an anisotropic fluid and the modified shear scalar dynamics led by it.

In this work, on the other hand, relying on energy-momentum squared gravity (EMSG), we look for a new possibility of that the stiff fluid-like shear scalar is retained (i.e., no anisotropic stresses employed) but its contribution to $H(z)$ is compensated by CDM, so that, for instance, the CMB quadrupole temperature fluctuation can be manipulated with giving rise to no deviation, on average, from either the standard $\Lambda$CDM model or the standard BBN. From the Einstein-Hilbert action of GR, it is possible to design a generalization involving nonlinear matter terms, by adding some analytic functions of a new scalar $T^{2}=T_{\mu \nu }T^{\mu \nu }$ formed from the energy-momentum tensor (EMT), $T_{\mu \nu }$, of the matter fields \cite{Katirci:2014sti}. Such generalizations result in new contributions by the usual matter fields to the right-hand side of the Einstein field equations without invoking new forms of matter and lead in general to nonconservation of the matter fields. The EMSG of the form $f(T^{2})=\alpha T^{2}$ (with $\alpha$ being a real constant), which considers simply the linear contributions of the new scalar, has been studied in various contexts in \cite{Roshan:2016mbt,Akarsu:2017ohj,Board:2017ign,Akarsu:2018zxl,Nari:2018aqs,Faria:2019ejh,Bahamonde:2019urw,Chen:2019dip,Barbar:2019rfn,Kazemi:2020hep,Rudra:2020rhs,Singh:2020bdv,Nazari:2020gnu}. The EMSG of this form is unique in that the dust in this case satisfies the conservation of the EMT and yet its linear (usual) contribution, $\rho _{\mathrm{m}}$, to the $H(z)$ is accompanied by its quadratic (new) contribution, $\alpha \rho _{\mathrm{m}}^{2}$, which mimics stiff fluid as exactly like the shear scalar does too. It is noteworthy that such an additional quadratic contribution of the matter energy density is reminiscent of the braneworld scenarios \cite{Brax:2003fv} for $\alpha>0$ and the loop quantum gravity \cite{Ashtekar:2011ni} for $\alpha<0$.

The observational upper limits on the present-day density parameter of a stiff fluid-like term included in the standard $\Lambda$CDM Friedmann equation can be adopted from \cite{Akarsu:2019pwn}; it is $\sim10^{-15}$ from the latest cosmological data (viz., joint CMB and BAO dataset), and $\sim10^{-23}$ from BBN. Thus, in both extensions of the standard $\Lambda$CDM model---i.e., in either its simplest anisotropic extension or its extension via the CDM coupled to the gravity in accordance with the EMSG of the form $f(T^{2})=\alpha T^{2}$---, the stiff fluid-like term involving in the Friedmann equation should today be very small (viz., the corresponding present-day density parameter should be less than $10^{-23}$) not to spoil the successful description of the Universe all the way to the BBN era. This might give the impression that such extensions to the standard $\Lambda$CDM model are permitted to be relevant only to the dynamics of the Universe well before the BBN. In what follows in the paper, we will discuss and show that this is not the case, particularly, when these two extensions are simultaneously employed. We proceed with constructing a generalization of the standard $\Lambda$CDM model, wherein we simultaneously replace the spatially flat RW metric with its simplest anisotropic generalization (LRS Bianchi I), and couple the CDM to the gravity in accordance with the EMSG of the form $f(T^2)\propto T^2$, while all other sources exist in the standard model of particle physics couple as usual in accordance with GR. Then we will focus on that these two modifications can mutually cancel out owing to the possibility of $\alpha<0$ (for which the new contributions of the CDM will resemble a stiff fluid with a negative energy density), viz., the shear scalar can be screened completely, and reproduce mathematically exactly the same Friedmann equation of the standard $\Lambda$CDM model. This allows us to get around the BBN limits on the anisotropic expansion, and thereby provides us an opportunity to manipulate the CMB quadrupole temperature fluctuation at the desired amount through a slightly anisotropic expansion in the late Universe. We further discuss the consequences of this model on the very early times and far future of the Universe, and finally briefly that such constructions may even be considered in the context of a major problem of the current cosmology, the so-called $H_0$ problem \cite{DiValentino:2020zio}.
 
\section{Model}
\label{sec:model}

We begin with the action constructed by the inclusion of the term $f\left(T_{\mu\nu}T^{\mu\nu},\mathcal{L}_{\rm m}\right)$ in the usual Einstein-Hilbert action with a bare cosmological constant $\Lambda$ as follows \cite{Akarsu:2018aro};
\begin{equation}
S=\int {\rm d}^4x \sqrt{-g}\,\left[\frac{1}{2\kappa}\left(R-2\Lambda \right)+f\left(T_{\mu\nu}T^{\mu\nu}, \mathcal{L}_{\rm m}\right)\right],
\label{action}
\end{equation}
where $\kappa$ is Newton's constant scaled by a factor of $8\pi$ (henceforth $\kappa=1$), $R$ is the scalar curvature, $g$ is the determinant of the metric $g_{\mu\nu}$, $\mathcal{L}_{\rm m}$ is the Lagrangian density corresponding to the matter field described by the energy-momentum tensor $T_{\mu\nu}$, and the units have been used such that $c=1$. We retain $\Lambda$ in accordance with the Lovelock's theorem stating that it arises as a constant of nature like $\kappa$ \cite{Lovelock:1971yv,Lovelock:1972vz}. In the usual fashion, we vary the action \eqref{action} with respect to the inverse metric $g^{\mu\nu}$ as
\begin{equation}
\begin{aligned}
\delta S=\int {\rm d}^4 x \sqrt{-g} \bigg[&\frac{1}{2}\delta R+\frac{\partial f}{\partial(T_{\alpha\beta}T^{\alpha\beta})}\frac{\delta(T_{\sigma\epsilon}T^{\sigma\epsilon})}{\delta g^{\mu\nu}}\delta g^{\mu\nu}\\
&+\frac{\partial f}{\partial \mathcal{L}_{\rm m}}\frac{\delta \mathcal{L}_{\rm m}}{\delta g^{\mu\nu}}\delta g^{\mu\nu}-\frac{1}{2}g_{\mu\nu}\delta g^{\mu\nu} \\
&\times \bigg\{\frac{1}{2}\left(R-2\Lambda\right)+f\left(T_{\sigma\epsilon}T^{\sigma\epsilon}, \mathcal{L}_{\rm m}\right)\bigg\}\bigg],
  \end{aligned} 
     \end{equation}
and define the EMT of the matter field as
  \begin{align}
  \label{tmunudef}
 T_{\mu\nu}=-\frac{2}{\sqrt{-g}}\frac{\delta(\sqrt{-g}\mathcal{L}_{\rm m})}{\delta g^{\mu\nu}}=g_{\mu\nu}\mathcal{L}_{\rm m}-2\frac{\partial \mathcal{L}_{\rm m}}{\partial g^{\mu\nu}},
 \end{align}
for which we assumed that $\mathcal{L}_{\rm m}$ depends only on the metric tensor components and not on its derivatives.

We proceed with the most straightforward example of the EMSG, which considers the linear contribution of the new scalar $T^{2}=T_{\mu \nu }T^{\mu \nu }$ in the action \eqref{action}, described by
\begin{equation}
\label{function}
f(T_{\mu\nu}T^{\mu\nu},\mathcal{L}_{\rm m})=\sum_i  \big(\alpha_i T_{\mu\nu}^{(i)}T^{\mu\nu}_{(i)}+\mathcal{L}_{\rm m}^{(i)}\big),
\end{equation} 
where, $i$ denoting the $i$th matter field (fluid), the summation over index $i$ is used for simplicity's sake as it avoids the cross-terms involving the product of the energy densities of different fluids in the field equations, and $\alpha_i$'s are constants that determine the coupling strength of the EMSG modifications to gravity for the $i$th fluid (cf. \cite{Akarsu:2018aro}). The action we proceed with is thus specified as follows;
\begin{equation}
\begin{aligned}
S=\int {\rm d}^4x \sqrt{-g}\,\bigg[\frac{R}{2}-\Lambda+\sum_i \big(\alpha_i T_{\mu\nu}^{(i)}T^{\mu\nu}_{(i)}+\mathcal{L}_{\rm m}^{(i)}\big)\bigg],
\label{eq:action}
\end{aligned}
\end{equation}
from which the modified Einstein field equations read
\begin{equation}
 \label{fieldeqn}
G_{\mu\nu}+\Lambda g_{\mu\nu}
=\sum_i T_{\mu\nu}^{(i)}+\sum_i\alpha_i \big(T_{\sigma\epsilon}^{(i)}T^{\sigma\epsilon}_{(i)} g_{\mu\nu}-2\,\Xi_{\mu\nu}^{(i)}\big).
\end{equation}
Here $G_{\mu\nu}=R_{\mu\nu}-\frac{1}{2}Rg_{\mu\nu}$ is the Einstein tensor and the new tensor is defined as
\begin{equation}
\begin{aligned}
\Xi_{\mu\nu}^{(i)}=&-2\mathcal{L}_{\rm m}^{(i)}\left(T_{\mu\nu}^{(i)}-\frac{1}{2}g_{\mu\nu}\mathcal{T}^{(i)}\right)-\mathcal{T}^{(i)}T_{\mu\nu}^{(i)} \\
&+2T^{\gamma(i)}_{\mu}T_{\nu\gamma}^{(i)}-4T^{\sigma\epsilon}_{(i)}\frac{\partial^2 \mathcal{L}_{{\rm m}}^{(i)}}{\partial g^{\mu\nu} \partial g^{\sigma\epsilon}},
\label{theta0}
\end{aligned}
\end{equation}
where $\mathcal{T}^{(i)}$ is the trace of the EMT of the $i$th fluid, $T_{\mu\nu}^{(i)}$, and the last term  vanishes as the EMT \eqref{tmunudef} does not include the second variation of $\mathcal{L}_{\rm m}^{(i)}$. We see, from \eqref{fieldeqn}, that the covariant divergence of the total EMT reads
\begin{equation}
\label{covdertmunu}
\nabla^{\mu} \sum_i T_{\mu\nu}^{(i)}=
-\nabla^{\mu}\sum_i \alpha_i \big(T_{\sigma\epsilon}^{(i)}T^{\sigma\epsilon}_{(i)} g_{\mu\nu}
-2\:\Xi_{\mu\nu}^{(i)}\big),
\end{equation}
which implies, unless $\alpha_i=0$ (GR), the total EMT is not conserved in general. We consider $\mathcal{L}_{\rm m}^{(i)}=p_i$ for the definition of the matter Lagrangian density that leads to the EMT of the form $T_{\mu\nu}^{(i)}=(\rho_{i}+p_i)u_{\mu}u_{\nu}+p_i g_{\mu\nu}$
(where $\rho_i$ and $p_i$ are, respectively, the energy density and the thermodynamic pressure of the $i{\rm th}$ fluid and $u_{\mu}$ is the four-velocity satisfying $u_{\mu}u^{\mu}=-1$ and $\nabla_{\nu}u^{\mu}u_{\mu}=0$) describing an isotropic perfect fluid form of matter field \cite{Bertolami:2008ab,Faraoni:2009rk}. Using this for barotropic equation of states as $w_i=\frac{p_i}{\rho_i}={\rm constant}$, we obtain
\begin{equation}
\label{thetafrw}
\begin{aligned}
T_{\mu\nu}^{(i)}T^{\mu\nu}_{(i)}=&\,\rho_i^2(3w_i^2+1),\\
\Xi_{\mu\nu}^{(i)}=&-\rho_i^2(3w_i+1)(w_i+1)u_{\mu}u_{\nu}.
\end{aligned}
\end{equation}
Thus, the covariant divergence of the total EMT \eqref{covdertmunu} reads
\begin{equation}
\begin{aligned}
\label{noncons}
\sum_i \left[\dot \rho_i+ \Theta(1+w_i)\rho_i\right]=
\sum_i \alpha_i\frac{2\Theta w_i(1+w_i)(5+3w_i)\rho_i^2}{1+2\alpha_i(1+8 w_i+3w_i^2)\rho_i},
\end{aligned}
\end{equation}
where $\Theta=\mbox{D}^{\mu}u_{\mu}$ is the volume expansion rate and a dot denotes derivative with respect to the comoving proper time $t$. Note that, unless $\alpha_i=0$ (GR), the local conservation of the total EMT is recovered only for $w_i={0,-1,-\frac{5}{3}}$.

We consider the locally rotationally symmetric (LRS) Bianchi I metric, the simplest anisotropic extension of the spatially flat RW metric,
\begin{equation}
\label{lrsbianchi}
{\rm d}s^2=-{\rm d}t^2+a^2(t)\,{\rm d}x^2+b^2(t)\,({\rm d}y^2+{\rm d}z^2),
\end{equation}  
where $\{a(t),b(t),b(t)\}$ are the directional scale factors along the principal axes $\{x,y,z\}$  \cite{Collins:1972tf,GEllisBook,Ellis:1998ct}. The corresponding average expansion scale factor reads $s(t)=(ab^2)^\frac{1}{3}$, and from which the average Hubble parameter $H=\frac{\Theta}{3}\equiv\frac{\dot{s}}{s}=\frac{1}{3}(H_a+2 H_b)$, where $H_a=\frac{\dot{a}}{a}$ and $H_b=\frac{\dot{b}}{b}$ are the directional Hubble parameters along the $x$- and $y$- (or $z$-) axes, respectively. And, we consider the usual cosmological fluids: CDM (c) and baryons (b) described by $w_{\rm c}=w_{\rm b}=0$, and radiation (photon $\gamma$ and neutrinos $\nu$) (r) described by $w_{\rm r}=\frac{1}{3}$. However, we suppose the CDM arbitrarily couples to gravity in accordance with the EMSG (i.e., $\alpha_{\rm c}$ is not necessarily null), while the particles present in the Standard Model (SM) of particle physics (b, $\gamma$, and three types of $\nu$) couple to gravity in the same way as in the GR (i.e., for these $\alpha_{\rm r}=\alpha_{\rm b}=0$). Consequently, calculating the relevant tensors given in \eqref{thetafrw}, and using \eqref{fieldeqn}, we reach the following set of modified Einstein field equations:
\begin{align}
2 H_a H_b+H_b^2-\Lambda=\rho_{\rm b}+\rho_{\rm c}+\alpha_{\rm c}\, \rho_{\rm c}^2+\rho_{\rm r},   \label{eq0:rho}\\
-2 \dot{H_b}-3H_b^2+\Lambda= \alpha_{\rm c}\, \rho_{\rm c}^2+\frac{\rho_{\rm r}}{3},  \label{eq0:px}    \\
-\dot{H_a}-\dot{H_b}-H_a^2-H_b^2-H_a H_b+\Lambda= \alpha_{\rm c}\, \rho_{\rm c}^2+\frac{\rho_{\rm r}}{3}. \label{eq0:py}
\end{align}

This set of equations can alternatively be written in terms of the average expansion rate $H(z)$ and the shear scalar $\sigma^2$ (which is defined, to quantify the anisotropic expansion, as $\sigma^2=\frac{1}{2} \sigma_{\alpha\beta} \sigma^{\alpha\beta}$, where $\sigma_{\alpha\beta}=\frac{1}{2}(u_{\mu;\nu}+u_{\nu;\mu}) h^{\mu}_{\:\alpha} h^{\nu}_{\:\beta}-\frac{1}{3} u^{\mu}_{\:;\mu} h_{\alpha\beta}$ is the shear tensor with $h_{\mu\nu}=g_{\mu\nu}+u_{\mu}u_{\nu}$ being the projection tensor \cite{Ellis:1998ct}). Accordingly, as $\sigma^2=\frac{1}{3}\left(H_{a}-H_{b}\right)^2$ for the LRS Bianchi I metric \eqref{lrsbianchi}, we reach
\begin{align}
 3 H^2-\sigma^2-\Lambda&=\rho_{\rm b}+\rho_{\rm c}+\alpha_{\rm c}\, \rho_{\rm c}^2+\rho_{\rm r}, \label{eq:rho2r}\\
-2 \dot{H}-3H^2-\sigma^2+\Lambda&=\alpha_{\rm c}\, \rho_{\rm c}^2+\frac{\rho_{\rm r}}{3}, \label{eq:pav}\\
\dot\sigma+3H\sigma&=0,
\label{eq:shearprop}
\end{align}
which are the energy density \eqref{eq:rho2r}, average pressure \eqref{eq:pav} and shear propagation \eqref{eq:shearprop} equations, respectively. It is reasonable to assume that, on cosmological scales, these matter fields are interacting only gravitationally, which leads to the separation of \eqref{noncons} into the different pieces for each one. We notice that, despite the fact that CDM contributes to the field equations in a modified way, it satisfies the local conservation of the EMT [i.e., \eqref{noncons} vanishes] and scales as usual as $\rho_{\rm c}=\rho_{\rm c0} s^{-3}$. Since the radiation and baryons couple to gravity as in the GR, these also scale as usual as $\rho_{\rm r}=\rho_{\rm r0} s^{-4}$ and $\rho_{\rm b}=\rho_{\rm b0} s^{-3}$. The shear propagation equation \eqref{eq:shearprop} dictates that the shear scalar scales as $\sigma^2=\sigma_0^2 s^{-6}$. Here, throughout the paper as well, a subscript 0 attached to any quantity denotes its present-day ($s=1$) value. Consequently, we reach the following modified Friedmann equation for our model:
\begin{equation}
\begin{aligned}
\label{exprate}
\frac{H^2}{H_0^2}=\;&\Omega_{\Lambda0}+\Omega_{\rm b0} s^{-3}+\Omega_{\rm r0} s^{-4}\\
&+\Omega_{\rm c0}\left(s^{-3}+\alpha'_{\rm c} s^{-6}\right)+\Omega_{\sigma0} s^{-6}, 
\end{aligned}
\end{equation}
where $\Omega_{\Lambda0}+\Omega_{\rm b0}+\Omega_{\rm r0}+\Omega_{\rm c0}(1+\alpha'_{\rm c})+\Omega_{\sigma0}=1$ with $\alpha'_{\rm c}\equiv\alpha_{\rm c}\,\rho_{\rm c0}$.  Here $\Omega_{i0}=\frac{\rho_{i0}}{3 H_0^2}$ are the present-day density parameters of the $i$th matter field, while $\Omega_{\Lambda0}=\frac{\Lambda}{3 H_0^2}$ and $\Omega_{\sigma0}=\frac{\sigma_0^2}{3 H_0^2}$ are those corresponding to $\Lambda$ and $\sigma^2$.

\section{$\Lambda$CDM with hidden anisotropic expansion}
\label{sec:cult}
Our model presents a mechanism for screening the shear scalar, which can even lead to the standard $\Lambda$CDM Friedmann equation in spite of anisotropic expansion: viz., collecting the like terms in \eqref{exprate} together we obtain
\begin{equation}
\begin{aligned}
\frac{H^2}{H_0^2}=&\; \Omega_{\Lambda0}+(\Omega_{\rm b0}+\Omega_{\rm c0}) s^{-3}+\Omega_{\rm r0} s^{-4}\\
&+\left(\Omega_{\sigma0}+\alpha'_{\rm c}\,\Omega_{\rm c0}\right) s^{-6},
\label{hubblep}
\end{aligned}
\end{equation}
wherein $\alpha'_{\rm c}\,\Omega_{\rm c0} s^{-6}$ (the quadratic contribution of the CDM energy density due to the EMSG) for $\alpha'_{\rm c}<0$ perpetually screens $\Omega_{\sigma0} s^{-6}$ (the contribution of the shear scalar), and the particular setting
\begin{equation}
\label{condition}
    \alpha'_{\rm c}=-\frac{\Omega_{\sigma0}}{\Omega_{\rm c0}}
\end{equation}
even hides it and leads to the Friedmann equation
\begin{align} \label{eq:lcdm}
\frac{H^2}{H_0^2}=(\Omega_{\rm b0}+\Omega_{\rm c0}) s^{-3}+\Omega_{\rm r0} s^{-4} 
+\Omega_{\Lambda0},
\end{align}
which is mathematically exactly the same with that of the standard $\Lambda$CDM model. Physically, on the other hand, the $H(z)$ here is the average expansion rate, and the expansion rates along the principal axes, viz., $H_a$ and $H_b$, need not necessarily be the same. This screening mechanism can be supposed to be working since then the times much before the BBN, as the CDM production is typically expected to occur much earlier than the BBN takes place---e.g., if CDM could be described by weakly interacting massive particles (WIMPs), starting from the energy scale $\sim0.1\,\rm TeV$ corresponding to the time (redshift) scale $\sim 10^{-10}\,\rm s$ ($z\sim10
^{15}$), whereas these are $\sim0.1\,\rm MeV$ and $\sim 100\,\rm s$ ($z\sim10^9$) for the standard BBN \cite{mukhanovbook}.

The model-independent upper limits on the present-day anisotropic expansion in terms of $\Omega_{\sigma 0}$ is of the order of $\mathcal{O}(10^{-3})$, e.g., from type Ia Supernovae \cite{Campanelli:2010zx,Wang:2017ezt} (see also \cite{Jimenez:2014jma,Soltis:2019ryf,Zhao:2019azy,Hu:2020mzd}). This, within the simplest anisotropic (i.e., Bianchi I) generalization of the standard $\Lambda$CDM ($\alpha'_{\rm c}=0$), implies the domination of the shear scalar at $z\sim 10$ and hence the spoilt of the successful description of the earlier ($z\gtrsim 10$) Universe. Indeed, while the constraint on a stiff fluid-like term ($\rho_{\rm s}=\rho_{\rm s0} s^{-6}$, likewise the shear scalar) on top of the standard $\Lambda$CDM model is $\Omega_{\rm s0}\lesssim 10^{-3}$ from the combined $H(z)$ and Pantheon data set (relevant to $z\lesssim 2.4$), it is tightened to $\Omega_{\rm s0}\lesssim 10^{-15}$ when the combined BAO and CMB (relevant to $z\sim1100$) data set also is included, and $\Omega_{\rm s0}\lesssim 10^{-23}$ upon demanding no significant deviation from the standard BBN (relevant to $z\sim10^9$) \cite{Akarsu:2019pwn}. All these can straightforwardly be adopted to our model upon defining $\Omega_{\rm s0}=\Omega_{\sigma0}+\alpha'_{\rm c}\,\Omega_{\rm c0}$ in \eqref{hubblep}. And our model, thus, can simultaneously accommodate the constraints $\Omega_{\rm s0}\lesssim10^{-23}$ (even $\Omega_{\rm s0}=0$) and $\Omega_{\sigma 0}\lesssim10^{-3}$, by means of the screening term $\alpha'_{\rm c}\,\Omega_{\rm c0}$ for suitably chosen values of $\alpha'_{\rm c}$. However, as the shear scalar still scales as $\sigma^2\propto s^{-6}$, the typical upper limit $\Omega_{\sigma 0} \sim 10^{-20}$ derived from the observed CMB quadrupole temperature fluctuation ($\Delta T/T \sim 10^{-5}$) setting an upper limit at the same order of magnitude on the anisotropy at the recombination era ($\sqrt{\Omega_{\sigma}
^{\rm rec}} \sim 10^{-5}$ at $z_{\rm rec}\sim10^3$) still applies \cite{Martinez95,Bunn:1996ut,Kogut:1997az,Saadeh:2016sak}. Consequently, one can think of manipulating the CMB quadrupole temperature via anisotropic expansion consistent with $\Omega_{\sigma 0} \sim 10^{-20}$ while retaining exactly the same expansion history for the comoving volume element of the Universe as that of the standard $\Lambda$CDM  all the way to the time (redshift) scale of $\sim 10^{-10}\,\rm s$ ($z\sim10
^{15}$), which can be promising, for instance, to address the so-called ``quadrupole temperature problem" \cite{Bennett11,Ade:2013kta,Schwarz:2015cma,Akrami:2019bkn}.

\section{Manipulating CMB Quadrupole Temperature Fluctuation}
\label{sec:qmanipulation}
As anisotropic expansion implies different evolution of the temperature of the free streaming photons for the different expansion factors in three orthogonal axes, it can be used for manipulating the quadrupole (multipole $\ell=2$ corresponding to the angular scale $\theta=\pi/2$) power spectrum of temperature fluctuations in the CMB, $\Delta T$, with no consequences on the higher multipoles. The evolution of the photon temperature along the $x$-axis and $y$-axis (or $z$-axis) is given by $T_{x} =T_0\frac{a_{0}}{a}=T_0 e^{-\int H_a {\rm d} t}$ and $T_{y}=T_0\frac{b_{0}}{b}=T_0 e^{-\int H_b {\rm d} t}$, where $T_{0}=2.7255\pm 0.0006\,{\rm K}$ \cite{Fixsen09} is the present-day CMB monopole temperature \cite{Barrow:1985tda,Barrow:1997sy}.
Accordingly, the difference between the photon temperatures along the $y$- and $x$-axes since the recombination ($z=z_{\rm rec}$) to the present time ($z=0$) due to the anisotropic expansion, $\Delta T_{\sigma}\equiv T_y-T_x$, reads
\begin{equation}
\begin{aligned}
\Delta T_{\sigma}&=T_0\int_{t_{\rm rec}}^{t_0}(H_a-H_b){\rm d}t =T_0 \int_{t_{\rm rec}}^{t_0}\sqrt{3} \sigma {\rm d}t\\
&=3T_0\sqrt{\Omega_{\sigma0}}\int_0^{z_{\rm rec}}\frac{H_0(1+z)^2}{H}{\rm d}z
\label{deltaT}
\end{aligned}
\end{equation}
for small anisotropies (so $e^{-\int H_a {\rm d} t}\simeq 1- \int H_a {\rm d} t$ etc). We use ${\rm d}t=-\frac{{\rm d}z}{H(1+z)}$ with $z=\frac{1}{s}-1$ being the average redshift defined from the average expansion scale factor and assume the CMB was last scattered at the recombination redshift (epoch) $z_{\rm rec}$ ($t_{\rm rec}$). Thus, under the condition \eqref{condition} retaining exactly the same expansion history for the comoving volume element with that of the standard $\Lambda$CDM, we can have change in $\Delta T$ up to \begin{equation}
\begin{aligned}
\Delta T_{\sigma}=3T_0\sqrt{-\alpha'_{\rm c}\,\Omega_{\rm c0}}\int_0^{z_{\rm rec}}\frac{H_0(1+z)^2}{H}{\rm d}z
\label{deltaTEMSG}
\end{aligned}
\end{equation}
on top of the best fit standard $\Lambda$CDM model predicted value $\Delta T_{\rm std}\approx 34\, \mu K$ ($\Delta T_{\rm std+var} \approx 28 \,\mu K$  when the cosmic variance is included) \cite{Dodelson03,Campanelli:2006vb}, and bring it to the observed value by the Planck satellite $\Delta T_{\rm PLK} \approx 14 \,\mu K$ \cite{Ade:2013kta}. Namely, we can make use of the observational best fit values from the recent Planck release \cite{Aghanim:2018eyx}, $\Omega_{\rm b0}=0.049$ and $\Omega_{\rm c0}=0.264$, along with the recombination redshift $z_{\rm rec}\approx1090$ and the present-day radiation density parameter $\Omega_{\rm r0}\approx10^{-4}$. Then, if we set $\Omega_{\sigma0}=4\times 10^{-21}$---corresponding to $\alpha'_{\rm c}= -1.52 \times10^{-20} \;(\alpha_{\rm c}=-1.8 \times10^{-11} \rm cm^{3}/erg$) from \eqref{condition}---we obtain $\Omega_{\sigma}
^{\rm rec}=1.23\times 10^{-11}$ along with that $\Delta T_{\sigma}= 20.5\;\mu \rm K$, which, provided that the orientation of the expansion anisotropy is set suitably, can reduce $\Delta T$ from $\Delta T_{\rm std}\approx 34 \mu \rm K$ predicted within the standard $\Lambda$CDM to the observed value $\Delta T_{\rm PLK} \approx 14 \,\mu K$.
 
For the radiation dominated era (for $z>z_{\rm eq}$, where $z_{\rm eq}=-1+(\rho_{\rm c0}+\rho_{\rm b0})/\rho_{\rm r0}$ is the matter-radiation equality redshift)---during which $\Lambda$ and the usual (linear) contribution of the CDM energy density are negligible but the shear scalar and the new (quadratic) contribution of the CDM energy density are subdominant---we can rewrite Eq.~\eqref{eq:rho2r} as $3H^2=\rho_{\rm r}+\rho_{\rm s}$, where $\rho_{\rm r}=\frac{\pi^2}{30}g_{*}T
^4$ and $\rho_{\rm s}=\alpha_{\rm c}\, \rho_{\rm c}^2+\sigma^2$, or, in a more useful form, as $3H^2=\frac{\pi^2}{30}\tilde{g}_{*}T^4$ with $\tilde{g}_{*}=(1-\Omega_{\rm s})^{-1}g_{*}$, the modified effective number of degrees of freedom, where $\Omega_{\rm s}=\Omega_{\sigma}\left(1+\alpha'_{\rm c}\frac{\Omega_{\rm c0}}{\Omega_{\sigma 0}}\right)$ and $g_{*}$ is the usual effective number of degrees of freedom counting the number of relativistic species determining the radiation energy density (cf. \cite{Barrow1976,Campanelli2011}). In the SM at $T=1\,\rm MeV$, $g_{*}= 5.5+\frac{7}{4}N_{\nu}$, where $N_{\nu}=3$ ($N_{\nu}=3.045$ when small corrections for nonequilibrium neutrino heating are included in the thermal evolution) is the effective number of (nearly) massless neutrino flavors \cite{Dodelson03}. $\tilde{g}_{*}$ is usually parametrized by $\Delta N_{\nu}=N_{\nu}-3$ (the deviation of $N_{\nu}$ from the SM value $N_{\nu}=3$) as $\tilde{g}_{*}=(1+\frac{7}{43}\Delta N_{\nu})g_{*}$. Consequently, at the time of freeze-out, viz., when the rate of the weak-interaction that interconverts neutrons and protons falls behind the Hubble expansion rate at $T_{\rm fr}\sim 1\,\rm MeV$, these two relations given above for $\tilde{g}_{*}$ imply that the stiff fluid-like term ($\rho_{\rm s}$) in our model can be regarded as a change in the total number of effectively massless degrees of freedom as
\begin{equation}
\label{Sandneutrino}
    \Omega_{\rm s}^{\rm fr}=\frac{7}{43}\Delta N_{\nu}
\end{equation}
for small $\Omega_{\rm s}^{\rm fr}$ values---so $(1-\Omega_{\rm s}^{\rm fr})^{-1}\simeq1+\Omega_{\rm s}^{\rm fr}$. This can then be translated into the density parameter of the stiff fluid-like term at the recombination through the relation
\begin{equation}
\Omega_{\rm s}^{\rm rec}=\Omega_{\rm s}^{\rm fr}(1+z_{\rm fr})^{-2}(1+z_{\rm eq})^{-1}(1+z_{\rm rec})^{3}
\end{equation}
(cf. \cite{Barrow:1997sy}). For the freeze-out redshift $z_{\rm fr}\sim 10^{9}$, consistent with the standard BBN, along with $z_{\rm eq}=3390$ and $z_{\rm rec}=1090$ from the best fit values of the standard $\Lambda$CDM in the recent Planck release \cite{Aghanim:2018eyx}, it turns out that $\Omega_{\rm s}^{\rm rec}=6.23\times 10^{-14} \Delta N_{\nu}$ (or $\Omega_{\rm s}^{\rm rec}=3.83\times 10^{-13} \Omega_{\rm s}^{\rm fr}$). Next, using $\Omega_{\rm b0}=0.049$, $\Omega_{\rm c0}=0.264$, and $\Omega_{\rm r0}\approx10^{-4}$ as well, we obtain $\Omega_{\rm s0}=3.24\times 10^{-10}\Omega_{\rm s}^{\rm rec}$ implying $\Omega_{\rm s0}=2.02\times 10^{-23} \Delta N_{\nu}$. All these, finally, lead to $\Omega_{\rm s}^{\rm fr}=0.163$, $\Omega_{\rm s}^{\rm rec}=6.23\times 10^{-14}$ and $\Omega_{\rm s0}=2.02\times 10^{-23}$ for $\Delta N_{\nu}=1$, and $\Omega_{\rm s}^{\rm fr}=0.05$, $\Omega_{\rm s}^{\rm rec}=1.87\times 10^{-14}$ and $\Omega_{\rm s0}=6.06\times 10^{-24}$ for the upper limit of $\Delta N_{\nu}=0.30$ from the recent Planck release \cite{Aghanim:2018eyx}. 

In the case of the straightforward Bianchi I extension of the standard $\Lambda$CDM ($\alpha'_{\rm c}=0$), these limits simply correspond to the limits on the shear scalar and hence, through \eqref{deltaT}, on $\Delta T_{\sigma}$ as well. Namely, now, we have $\Omega_{\sigma}^{\rm fr}=0.05$, $\Omega_{\sigma}^{\rm rec}=1.87\times 10^{-14}$, and $\Omega_{\sigma0}=6.06\times 10^{-24}$ leading to $\Delta T_{\sigma}= 0.82\;\mu \rm K$. Thus, in this case, the BBN restricts the possible manipulation upon the CMB quadrupole temperature fluctuation via the anisotropic expansion to insignificant values (viz., $\Delta T_{\sigma}\lesssim 1 \mu \rm K$). In our model, on the other hand, the limit $\Omega_{\rm s0}\lesssim 10^{-23}$ required by the BBN does not necessarily lead to $\Delta T_{\sigma}\lesssim 1 \mu \rm K$. It can still be satisfied when $\Omega_{\sigma0}\sim 10^{-21}$ (or  $\Omega_{\sigma}^{\rm rec}\sim 10^{-11}$), which leads to an amount of manipulation upon the CMB quadrupole temperature fluctuation on the same order of magnitude with its observed value, provided that the gravitational coupling of the CDM is augmented by the EMSG with $\alpha'_{\rm c}\sim-10^{-20}$. Moreover, under the condition \eqref{condition}, we reproduce exactly the same expansion history with that of the standard $\Lambda$CDM cosmology all the way to BBN era with an additional opportunity of manipulating the CMB quadrupole temperature fluctuation at desired values. Thus, our model provides us with opportunity to fine tune the CMB quadrupole temperature fluctuation (e.g., for addressing the so-called `quadrupole temperature problem') without leading to any other measurable alteration in the standard $\Lambda$CDM.

\section{Early and late dynamics}
\label{sec:fandp}

We have reached, by eliminating the terms scaling as $s^{-6}$ in \eqref{hubblep} via the condition $\Omega_{\rm s0}=0$ given in \eqref{condition}, exactly the same mathematical form of the Friedmann equation of the standard $\Lambda$CDM, where however physically, $H(z)$ is the average expansion rate and anisotropic expansion is allowed. This relies on the cooperation between the CDM coupled to gravity in accordance with the EMSG of the form $f(T^2)\propto T^2$ and the anisotropic expansion, and hence will be valid all the way to the CDM generation redshift $z_{\rm c}$. And, this redshift is typically considered to be much larger than the BBN redshift $z_{\rm BBN}\sim z_{\rm fr}$. Therefore, even if it is guaranteed that the average expansion rate of the Universe during BBN equals the one in the standard BBN (in spite of that $\Omega_{\sigma0}=4\times 10^{-21}$, which leads to $\Delta T_{\sigma}\approx 20.5\;\mu \rm K$ manipulation in the CMB quadrupole temperature fluctuation), for the times $z>z_{\rm c}$ (i.e., when CDM did not exist yet) the Universe is described by the general relativistic LRS Bianchi I cosmological model \cite{GEllisBook,Ellis:1998ct} in the presence of radiation (which approximates the LRS Kasner vacuum solution \cite{Stephani2003} with the increasing redshift). On the other hand, this opportunity of letting safely to $\Delta T_{\sigma}\approx 20.5\;\mu \rm K$ manipulation is in fact not subject to the condition $\Omega_{\rm s0}=0$ (which evades BBN limits), but $\big|\Omega_{\rm s0}\big|\lesssim 10^{-23}$ (corresponding to $\big|\Delta N_{\nu}\big|\lesssim 0.30$ in line with the limits given in the recent Planck release \cite{Aghanim:2018eyx}). Consideration of this slightly relaxed condition gives rise to several other possibilities for the dynamics of the early Universe for $z>z_{\rm BBN}$: (I) In the case of $0<\Omega_{\rm s0}\lesssim 10^{-23}$, as $z$ increases,
the stiff fluid-like term domination over radiation can develop at a redshift either smaller or larger than $z_{\rm c}$. And,  for $z>z_{\rm c}$, the Universe is described by the general relativistic LRS Bianchi I cosmological model in the presence of radiation. (II) In the case of $- 10^{-23}\lesssim\Omega_{\rm s0}<0$,
the stiff fluid-like term---which, in this case, yields negative energy density as $\alpha'_{\rm c}\Omega_{\rm c0}<-\Omega_{\sigma0}<0$---brings in the following three different scenarios: (a) As $z$ increases, the stiff fluid-like term slows down the increment of $H(z)$ in redshift, but $z=z_{\rm c}$ is reached before it starts to decrease $H(z)$ itself. And, for $z>z_{\rm c}$, the Universe is described by the general relativistic LRS Bianchi I cosmological model in the presence of radiation. (b) As $z$ increases, the stiff fluid-like term slows down the increment of $H(z)$ in redshift and then it starts to decrease $H(z)$ itself, but $z=z_{\rm c}$ is reached before $H(z)$ vanishes. And, for $z>z_{\rm c}$, the Universe is described by the general relativistic LRS Bianchi I cosmological model in the presence of radiation (so, $H(z)$ starts to increase with redshift once again). (c) As $z$ increases, the stiff fluid-like term slows down the increment of $H(z)$ in redshift and it eventually decreases $H(z)$ until it vanishes completely before $z=z_{\rm c}$ is reached. This is the most interesting one among the possible scenarios, as it implies that the CDM was never generated but was always there, and that the Universe started to expand from a nonzero volume.

As the Universe continues to expand in the future (when $-1\leq z<0$), both the deviation from GR (viz., the  quadratic contribution of the CDM energy density due to the EMSG) and the expansion anisotropy (viz., the shear scalar) keep on diluting faster than all the other terms that constitute the standard $\Lambda$CDM part in \eqref{hubblep}, namely, our model will asymptotically approach the usual standard $\Lambda$CDM model---i.e., the Universe isotropizes and the EMSG approaches the GR---and the de Sitter solution in the arbitrarily far future.

We have contented ourselves with just commenting on the very early ($z>z_{\rm BBN}$) and future ($z<0$) dynamics of the Universe, rather than presenting a comprehensive analysis. Yet, one may find it quite enlightening to see Ref. \cite{Chavanis:2014lra}---examines the cosmological model which includes stiff fluid source on top of the standard $\Lambda$CDM model---regarding, in particular, the evolution of the average expansion scale factor in our model, and Ref. \cite{Barrow78}---presents an investigation of anisotropic cosmologies in the presence of stiff fluid with a positive energy density (reminiscent of our model for $\alpha_{\rm c}>0$).

\section{Remarks and future perspectives}
\label{sec:conclusion}

The simplest anisotropic extension of the standard $\Lambda$CDM model---just extends the spatially flat RW metric to the Bianchi I---leads to a generalized Friedmann equation that brings in the average Hubble parameter $H(z)$ along with a shear scalar of the form $\sigma^2=\sigma^2_0 (1+z)^6$, i.e., mimicking stiff fluid with a positive energy density \cite{Akarsu:2019pwn}. Our model \eqref{exprate} then just adds into it another stiff fluid-like term, $\alpha_{\rm c} \rho_{\rm c}^2=\alpha'_{\rm c}\rho_{\rm c0}(1+z)^6$, the quadratic contribution of the CDM energy density scaled by the constant $\alpha'_{\rm c}$, which is not necessarily positive while determining the gravitational coupling strength of CDM in accordance with the EMSG of the form $f(T^2)\propto T^2$. The Bianchi I metric, however, is atypical in that it brings in no restoring forcelike term in the shear propagation equation [cf. \eqref{eq:shearprop}], whereas one set of such terms come, in more complicated anisotropic metrics, from anisotropic spatial curvature of the metric itself \cite{GEllisBook,Ellis:1998ct,Barrow:1997sy}. This implies that the stiff fluid-like shear scalar is not generic, even for the general relativistic cosmologies in the presence of usual cosmological fluids (isotropic perfect fluids with no peculiar velocities) only. For instance, the Bianchi VII$_0$ metric \cite{GEllisBook,Ellis:1998ct}---the most general spatially homogeneous and flat anisotropic metric---yields, in addition to the simple expansion-rate anisotropies present in the Bianchi I, an anisotropic spatial curvature that resembles a traceless anisotropic fluid \cite{Barrow:1997sy}. It causes, in the general relativistic universes close to isotropy, the shear scalar to scale as $\sigma^2\propto (1+z)^5$ during the dust era and as $\sigma^2\propto \ln (z/z_{\rm fr})^{-2}(1+z)^4$ during the radiation era, hence the limit on its present-day density parameter from BBN to be weaker than that from CMB---in contrast to the situation in the Bianchi I---, and both of the limits to be weaker than the ones derived when the Bianchi I metric is considered \cite{Barrow:1997sy,Barrow1976}. It is conceivable that, if we switch to the Bianchi VII$_0$ metric in our model as well, it will cause the same shear scalar dynamics. For, contrarily to the modified theories of gravity (e.g., the scalar-tensor theories of gravity \cite{Madsen88,Pimentel89,Faraoni:2018qdr,Akarsu:2019pvi}) in general, the EMSG does not induce nonzero anisotropic stresses \cite{Board:2017ign}, and therefore leads to the same shear propagation equations [cf. \eqref{eq:shearprop}] with the usual ones derived in GR. Consequently, as the shear scalar in this case will grow slower than the stiff fluid-like contribution of CDM, it is no more possible to achieve the mutual cancellation of these two terms perpetually and write \eqref{eq:lcdm}. Thus, if we reconsider our model by switching to the Bianchi VII$_0$ metric, we expect the strongest limits upon the shear scalar to come from CMB, and the ones on the stiff fluid-like contribution of CDM to come from BBN (viz., as in this case we can write $\Omega_{\rm s0}=\alpha'_{\rm c}\,\Omega_{\rm c0}$, it will be necessary to satisfy $|\alpha'_{\rm c}\,\Omega_{\rm c0}|\lesssim 10^{-23}$---implying $|\alpha'_{\rm c}|\,\lesssim 10^{-22}$ for $\Omega_{\rm c0}\sim 0.25$---corresponding to $\big|\Delta N_{\nu}\big|\lesssim 0.30$ in line with the limits given in the recent Planck release \cite{Aghanim:2018eyx}).

The discussion in the previous paragraph shows also that, to create a measurable change in the CMB quadrupole temperature fluctuation without spoiling the successes of the standard BBN, it is no more needed in the case of the Bianchi VII$_0$ to apply the mechanism of screening the shear scalar by the stiff fluid-like contribution of CDM. Indeed, it is well known that the strong limits upon the shear scalar (so the anisotropic expansion) are usually model-dependent and can be vastly weakened by promoting its simplest stiff fluid-like behavior to a more complex dynamical one by means of an anisotropic fluid (either an actual source or an effective source from a modified gravity theory) and/or a nontrivial anisotropic spatial curvature exits in more generic anisotropic metrics such as the Bianchi VII$_0$ \cite{Barrow:1997sy}. Our work distinguishes from such works as it studies a possibility of an alternative mechanism weakening the limits upon shear scalar through screening its contribution to $H(z)$ instead of modifying it. Namely, by counterbalancing the shear scalar term $\Omega_{\sigma0} (1+z)^{6}$ via the new term $\alpha'_{\rm c}\,\Omega_{\rm c0} (1+z)^{6}$ from the gravitational coupling of CDM in accordance with the EMSG of the form $f(T^2)\propto T^2$, we have evaded the limits upon the anisotropic expansion coming from the enhancing influence of the shear scalar on $H(z)$ (e.g., the limits from BBN), but kept on using the ones coming directly from the anisotropy in the expansion itself (e.g., the limits from the CMB quadrupole temperature fluctuations). This feature of our model would be more significant, if it turns out that there is one additional neutrino species beyond the three predicted by the Standard Model of particle physics (as, e.g., suggested for alleviating the so-called $H_0$ tension \cite{Carneiro:2018xwq}). For, it amounts to $\Omega_\sigma\sim 0.16$ during BBN and so leaves less room for the anisotropic expansion (see Sec. \ref{sec:qmanipulation}), but there can still be anisotropic expansion large enough to have a measurable effect in the CMB radiation, since we can still evade the BBN limits by compensating the contributions both from the shear scalar and additional neutrino species.

In our study, we have focused on the aspect of the model that the matter field coupled to the gravity in accordance with a suitably arranged EMSG setup can compensate for the enhancing influence of anisotropy on the average expansion rate of the Universe. Yet, through this model, we have learned also lessons on some other aspects of the cosmological models that employ EMSG. It would be useful to briefly mention the some that may give insight into the possible prospective works. In the literature to date, the EMSG of the form $f(T^2)\propto T^2$, as well as its power-law generalization $f(T^2)\propto T^{2\eta}$ with $\eta>\frac{1}{2}$ (known also as EMPG), has been mostly studied in the context of the early Universe dynamics and used, particularly, to avoid---replace with a nonsingular beginning/bounce---the initial big bang singularity \cite{Roshan:2016mbt,Akarsu:2017ohj,Board:2017ign,Akarsu:2018zxl,Bahamonde:2019urw,Barbar:2019rfn,Nazari:2020gnu}. For, the new contributions of the matter field to the Friedmann equation in these studies scale faster than the usual (linear) contributions; therefore, the earlier times the more effective these new contributions are. We, however, notice that all these studies consider spatially homogeneous and isotropic RW metric and then the inclusion of anisotropy can prevent such scenarios from happening. Namely, it is possible that, as we move backward in time, the shear scalar grows fast enough to dominate over the new contributions of the matter field before these could give rise to a nonsingular beginning/bounce and then the very early Universe will be best described by the usual anisotropic spacetime vacuum solutions of GR (e.g., by the Kasner vacuum solution). In a realistic description of the Universe one can suppose the observable Universe is almost-exactly isotropic but not exactly isotropic. Therefore, it is important to pick, among these scenarios developed under the RW metric assumption, the ones that can survive when anisotropy is included. One another lesson is that, the EMSG models that add, into the Friedmann equation, the new contributions of the matter fields scaling faster than the usual (linear) contributions do have consequences on not only the early universe but also the late Universe. The particular model we have studied here presents a good example of this, as it evades the BBN limits on the present-day expansion anisotropy of the Universe. And  a closer look reveals that, beyond the limited framework we have drawn in this work, it may have consequences on a major problem relevant to the present-day Universe, the so-called $H_0$ problem. The stiff fluid-like term for $\alpha_{\rm c}>0$ in our model can be regarded as an increment in the total number of effectively massless degrees of freedom [see Eqn. \eqref{Sandneutrino}], which has been considered as one of the possible solutions for the $H_0$ problem \cite{DiValentino:2020zio}. Finally, the study we have carried out here can be extended to more complicated constructions by considering more generic anisotropic metrics and/or functions of $f(T^2)$, albeit, most likely, one will need to compromise both the energy-momentum  conservation law and simplicity we have had in this particular setup here.

\begin{acknowledgements}
The authors thank to Nihan Kat{\i}rc{\i}, Suresh Kumar and Jorge Nore\~{n}a for discussions. \"{O}.A. acknowledges the support by the Turkish Academy of Sciences in scheme of the Outstanding Young Scientist Award  (T\"{U}BA-GEB\.{I}P). \"{O}.A. also acknowledges the support received from, and hospitality of the Abdus Salam International Centre for Theoretical Physics (ICTP), where most of this work was carried out. J.D.B. is supported by the Science and Technology Funding Council (STFC) of the UK.
\end{acknowledgements}

\end{document}